# Exploration quantum steering, nonlocality and entanglement of two-qubit *X*-state in structured reservoirs

Wen-Yang Sun, Dong Wang, Jia-Dong Shi and Liu Ye[*]

School of Physics & Material Science, Anhui University, Hefei 230601, China

**In this work, there are two parties, Alice on Earth and Bob on the satellite, which initially share an entangled state, and some open problems, which emerge during quantum steering that Alice remotely steers Bob, are investigated. Our analytical results indicate that all entangled pure states and maximally entangled evolution states (EESs) are steerable, and not every entangled evolution state is steerable and some steerable states are only locally correlated. Besides, quantum steering from Alice to Bob experiences a "sudden death" with increasing decoherence strength. However, shortly after that, quantum steering experiences a recovery with the increase of decoherence strength in bit flip (BF) and phase flip (PF) channels. Interestingly, while they initially share an entangled pure state, all EESs are steerable and obey Bell nonlocality in PF and phase damping channels. In BF channels, all steerable states can violate Bell-CHSH inequality, but some EESs are unable to be employed to realize steering. However, when they initially share an entangled mixed state, the outcome is different from that of the pure state. Furthermore, the steerability of entangled mixed states is weaker than that of entangled pure states. Thereby, decoherence can induce the degradation of quantum steering, and the steerability of state is associated with the interaction between quantum systems and reservoirs.**

Quantum entanglement has been a topic of great interest ever since the pioneering work was presented by Einstein *et al.*[1] in 1935. It is defined as the nonseparability of quantum states[2-6], and is one of the most important resources in quantum information processing. Notably, correlations arising from local measurements performed on separated entangled systems can exhibit nonlocal correlations[7, 8]. In particular, the observed statistics cannot be reproduced using a local hidden variable model, as witnessed by violation of a Bell inequality[2, 3].

Originally, the phenomenon of Einstein-Podolsky-Rosen (EPR) steering (or quantum steering)

---

[*] Corresponding author: yeliu@ahu.edu.cn

was introduced by Schrödinger in 1935 to analyze the EPR-paradox[9, 10]. Later, some theoretical and experimental works concerning quantum steering have been achieved[11-21], and Wiseman *et al.*[22, 23] formulated steering in an operational way in conformity for a quantum information task. Recently, quantum steering was given an operational explanation as the distribution of entanglement by an untrusted party[22], which depends on the question of whether Alice can convince Bob when they share an entangled state, although the fact that Bob distrust Alice. Then, Alice performs her measurements (which are unknown to Bob) and informs him of the results. If the correlations between Bob's measurement results and those Alice reports cannot be explained by a local hidden state model (LHSM)[23] for Bob, then Bob will believe that they share an entangled state. Quantum steering is an intermediate form of quantum correlation between Bell nonlocality[2, 8] and entanglement[3] in modern quantum information theory. Furthermore, quantum steering can be detected via violating quantum steering inequality[24]. Derived for both continuous and discrete variable systems[25-29], such steering inequalities can be obtained employing entropic uncertainty[27, 30]. The significant steering criteria have been developed[31-37] to detect steering from different aspects. These criterions can also be used to guarantee one-way steering[15], namely, Alice can steer Bob, however Bob cannot steer Alice. And the one-way steering has been verified in some theoretical and experimental works[11-17].

Despite previous fruitful achievements, however, these investigations mentioned are limited to the exploration of quantum steering in an isolated system. In a realistic regime, quantum systems unavoidably suffer from decoherence or dissipation arising from the interaction between the systems and its external noises[38, 39]. Consequently, it is important to investigate quantum steering, nonlocality and entanglement under the influence of reservoirs (noisy channels), and establish whether the steerable state depends on reservoirs. As a matter of fact, there are a few authors to pay attention to address this problem[40-42]. In this work, some problems of that Alice can distantly steer Bob are investigated, and then we consider two different types of bipartite states (entangled pure state and entangled mixed state) as the initial states. Herein, we explore the performance of quantum steering, nonlocality and entanglement in the different reservoirs. Our analytical results indicate that: (i) All entangled pure states and maximally entangled evolution states are steerable. (ii) Not every entangled evolution state is steerable and some steerable states cannot violate Bell-CHSH inequality. (iii) Decoherence can rapidly induce the degradation of quantum steering, and the steerability of entangled pure states is stronger than that of entangled mixed states.

## Results

**Exploring the performance of quantum steering, entanglement and nonlocality of two-qubit *X*-state in the different reservoirs**

We assumed that there are two parties, Alice on Earth and Bob on the satellite, sharing a pair of entangled photons. Then we will elaborate the steering, nonlocality and entanglement in a physical case illustrated in Fig. 1 (a) as following: Alice prepares a pair of entangled photons and sends one to Bob. The photon *B* in the process of transmission inevitably suffers from the different noisy environments[43] (amplitude damping (AD), phase damping (PD), phase flip (PF) and bit flip (BF) channels). We will investigate the performance of quantum steering, nonlocality and entanglement for the evolution state described by a trace-preserving quantum operation $\varepsilon(\rho)$, which is given by $\varepsilon(\rho) = \sum_{i=0,1} \left( I^A \otimes E_i^B \right) \rho \left( I^A \otimes E_i^B \right)^\dagger$, where $\{E_i\}$ is the set of Kraus operators associated to a decohering process of a single qubit, with the trace-preserving condition reading[44] $\sum_i E_i^\dagger E_i = I$. Then, we provide lists of Kraus operators for varieties of quantum channels considered in Table. 1. Here, we define that the entangled evolution states (EESs) are damped states, which the subsystem *B* of the initial bipartite state suffers from the quantum noisy channels. We will consider two different types of initial states, entangled pure state and entangled mixed state:

| Channels | Kraus operators |
|---|---|
| PF | $E_0 = \sqrt{p}I, \ E_1 = \sqrt{1-p}\sigma_z$ |
| BF | $E_0 = \sqrt{p}I, \ E_1 = \sqrt{1-p}\sigma_x$ |
| AD | $E_0 = \begin{pmatrix} 1 & 0 \\ 0 & \sqrt{1-d} \end{pmatrix}, \ E_1 = \begin{pmatrix} 0 & \sqrt{d} \\ 0 & 0 \end{pmatrix}$ |
| PD | $E_0 = \begin{pmatrix} 1 & 0 \\ 0 & \sqrt{1-d} \end{pmatrix}, \ E_1 = \begin{pmatrix} 0 & 0 \\ 0 & \sqrt{d} \end{pmatrix}$ |

**Table. 1.** Kraus operators for the quantum channels: phase flip (PF), bit flip (BF), amplitude damping (AD) and phase damping (PD), where *d* and *p* are decoherence probabilities.

**Alice and Bob share an entangled pure state.** Assume that they have $|\varphi\rangle_{AB} = \cos\alpha|00\rangle + \sin\alpha|11\rangle$, $0 < \alpha < \pi/2$ and can also be expressed as

$$\rho = \cos^2\alpha |00\rangle\langle 00| + \sin^2\alpha |11\rangle\langle 11| + \cos\alpha\sin\alpha |00\rangle\langle 11| + \cos\alpha\sin\alpha |11\rangle\langle 00|. \quad (1)$$

Based on Eqs. (10) and (19) in the section of Methods, we can obtain its entanglement $C = \sin(2\alpha)$ and Bell-CHSH inequality $B = 2\sqrt{1 + \sin^2(2\alpha)}$, respectively. It is straightforward to insert Eq. (1) (via Eq. (17)) into Eq. (21) in the Methods, resulting in the analytical expression of entropic uncertainty relations (*EUR*) *steering inequality* for the density matrix $\rho$.

In order to better understand *EUR steering inequality*[27, 30] for a pair of arbitrary observables, we take advantage of the results of Walborn *et al.*[26]. The system is explained by a LHSM if and only if (iff) the joint measurement probability density can be expressed as[29-31]

$$\rho(x^A, x^B) = \int d\lambda \, \rho(\lambda)\rho(x^A|\lambda)\rho_q(x^B|\lambda), \quad (2)$$

where $\rho_q(x^B|\lambda)$ is the probability density (TPD) of measuring $\hat{x}^B$ to be $x^B$ given the details of preparation in the hidden variable $\lambda$. The subscript $q$ denotes that this is TPD arising from a single state. By applying the positivity of the continuous relative entropy[45] between any couple of probability distributions, Walborn *et al.*[26] argued that it is always the case for continuous observables (COs) in states allowing LHSM that $h(x^B|x^A) \geq \int d\lambda \, \rho(\lambda) h_q(x^B|\lambda)$, where $h_q(x^B|\lambda)$ is the continuous Shannon entropy caused by TPD. Then, it is straightforward to show (as Walborn et al. did) that any state allowing a LHSM in position and momentum must satisfy

$$h(x^B|x^A) + h(k^B|k^A) \geq \log(\pi e). \quad (3)$$

Note that here and throughout the paper the base of all logarithms is assumed to be 2. Subsequently, one notes that the same arguments used to develop LHSM constraints for COs can be employed to formulate LHSM constraints for discrete observables (DOs) as well[27]. Because the positivity of the relative entropy is a fact[45] for both continuous and discrete variables, one can derive the corresponding local hidden states constraint for DOs in the same way: $H(R^B|R^A) \geq \sum_\lambda P(\lambda) H_q(R^B|\lambda)$, where $H_q(R^B|\lambda)$ is the discrete Shannon entropy of $P_q(R^B|\lambda)$. Then, we immediately obtain a new entropic steering inequality for pairs of DOs[27]

$$H(R^B|R^A) + H(S^B|S^A) \geq \log(\Omega^B), \quad (4)$$

where $\Omega^B$ is the value $\Omega \equiv \min_{i,j}\left(1/\left|\langle R_i|S_j\rangle\right|^2\right)$, $\{|R_i\rangle\}$ and $\{|S_j\rangle\}$ are the eigenbases of observables $\hat{R}^B$ and $\hat{S}^B$ in the same $N$-dimensional Hilbert space, respectively. We must realize that for any *EUR*, even some relating more than two observables, there is a corresponding steering inequality[27]. Sánchez-Ruiz[46] developed *EUR* for complete sets of mutually unbiased observables $\{\hat{R}_i\}$, where $i=\{1, ..., N\}$. The $N$ is dimensionality of the system, it has been shown[47] that there are complete sets of $N+1$ mutually unbiased observables. We can obtain the *EUR* $\sum_i^{N+1}H(R_i) \geq (N/2)\log(N/2)+(1+N/2)\log(1+N/2)$ in even dimensional quantum systems. The *EUR* can be adapted into quantum steering inequality readily by substituting conditional entropies for marginal ones. In the same way as done to derive Eq. (4), we can obtain the *EUR steering inequality*[27]

$$\sum_k^{N+1} H\left(R_k^B|R_k^A\right) \geq (N/2)\log(N/2)+(1+N/2)\log(1+N/2), \quad (5)$$

where $H(B|A)=H(\rho_{AB})-H(\rho_A)$ is the conditional von Neumann entropy. In two dimensional quantum systems, in terms of Eq. (5), employing the Pauli *X*, *Y*, and *Z* measurements bases on each side, and then the *EUR steering inequality* can be read as[27]

$$H\left(\sigma_x^B|\sigma_x^A\right)+H\left(\sigma_y^B|\sigma_y^A\right)+H\left(\sigma_z^B|\sigma_z^A\right) \geq 2. \quad (6)$$

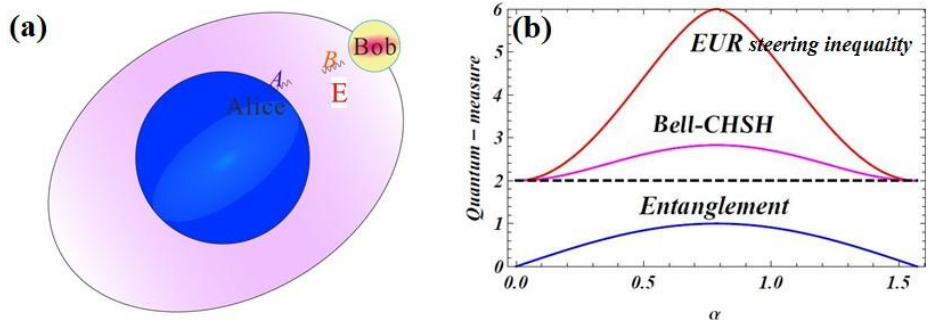

**Fig. 1.** (Color online) **(a)** Schematic diagram of systems: there are two parties shared an entangled state. Alice on the Earth and Bob on the satellite. If Alice can prepare a pair of entangled photons, Then, Alice sends one subsystem (photon *B*) of entangled photon to Bob. The photon *B* in the process of transmission inevitably suffers from the different noises. The red **E** denote noisy environment. **(b)** Varieties of quantum-measure (*EUR steering inequality*, Bell-CHSH inequality and entanglement) as function of the state parameters $\alpha$ when they initially share an entangled pure state.

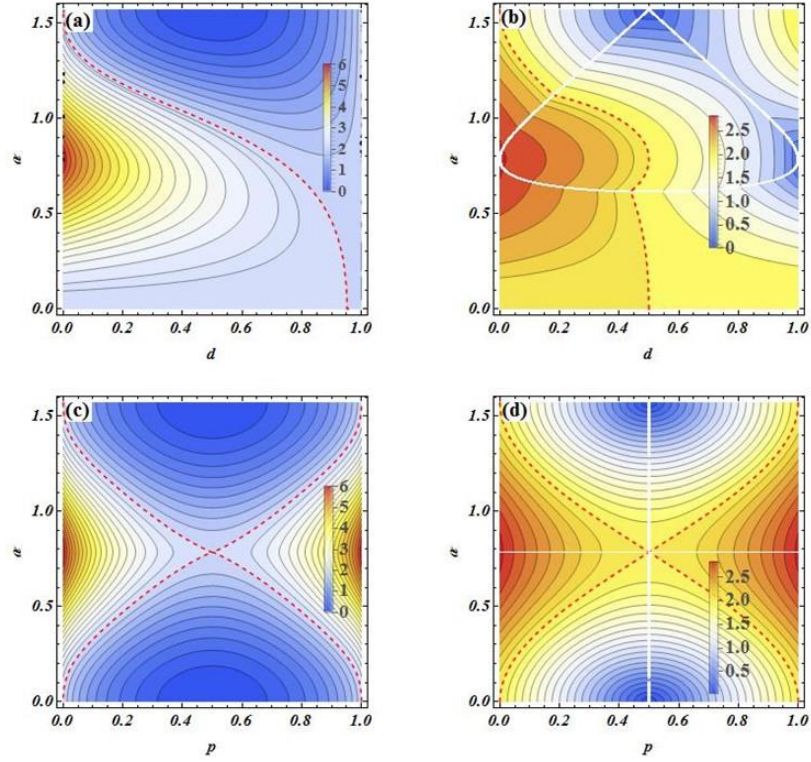

**Fig. 2.** (Color online) AD channel, contour plot of *EUR steering inequality* and Bell-CHSH inequality versus decoherence strength $d$ and states parameters $\alpha$ in **(a)** and **(b)**, respectively. The left side of red dotted line denotes the steering (shown in **(a)**) and Bell nonlocality (shown in **(b)**). For BF channel, contour plot of *EUR steering inequality* and Bell-CHSH inequality versus decoherence strength $p$ and states parameters $\alpha$ in **(c)** and **(d)**, respectively. The left and right sides of the *X*-form denote the steerable (shown in **(c)**) and Bell nonlocality (shown in **(d)**).

|  | AD | PD | PF | BF |
|---|---|---|---|---|
| $c_1$ | $\sqrt{1-d}\sin(2\alpha)$ | $\sqrt{1-d}\sin(2\alpha)$ | $(2p-1)\sin(2\alpha)$ | $\sin(2\alpha)$ |
| $c_2$ | $-\sqrt{1-d}\sin(2\alpha)$ | $-\sqrt{1-d}\sin(2\alpha)$ | $(1-2p)\sin(2\alpha)$ | $(1-2p)\sin(2\alpha)$ |
| $c_3$ | $1-d+d\cdot\cos(2\alpha)$ | 1 | 1 | $(2p-1)$ |
| $r$ | $\cos(2\alpha)$ | $\cos(2\alpha)$ | $\cos(2\alpha)$ | $\cos(2\alpha)$ |
| $s$ | $d-(d-1)\cos(2\alpha)$ | $\cos(2\alpha)$ | $\cos(2\alpha)$ | $(2p-1)\cos(2\alpha)$ |
| $\mu_1$ | $(\sqrt{1-d}\sin(2\alpha))^2$ | $(\sqrt{1-d}\sin(2\alpha))^2$ | $[(2p-1)\sin(2\alpha)]^2$ | $[\sin(2\alpha)]^2$ |
| $\mu_2$ | $(\sqrt{1-d}\sin(2\alpha))^2$ | $(\sqrt{1-d}\sin(2\alpha))^2$ | $[(2p-1)\sin(2\alpha)]^2$ | $[(2p-1)\sin(2\alpha)]^2$ |
| $\mu_3$ | $[1-d+d\cdot\cos(2\alpha)]^2$ | 1 | 1 | $(2p-1)^2$ |

**Table. 2.** The corresponding expression of each parameter of two-qubit EESs in Bloch decomposition and the parameters $\mu_1$, $\mu_2$, $\mu_3$ are given in the different channels when Alice and Bob initially share an entangled pure

state.

As shown in Fig. 1 (b), one can find that all entangled pure states are steerable and satisfy Bell nonlocality. Besides, the maximally entangled pure state ($\alpha = \pi/4$) is maximally steerable, say, Alice can perfectly remotely steer Bob. Next, let us investigate the performance of entanglement, nonlocality and quantum steering in the different quantum noisy channels. For simplicity, we will not write out detailed calculation process. The corresponding each parameter expression of two-qubit EESs in Bloch decomposition and the parameters $\mu_1$, $\mu_2$, $\mu_3$ are given in Table. 2.

To better understand the relationship between quantum steering and nonlocality in different noisy channels, we plot some graphs in Fig. 2. In AD channel, we can find that quantum steering decreases with the increase of decoherence strength, and until the state is unsteerable (i.e., the Bob does not trust Alice that they shared states are entangled) iff decoherence strength is very large (i.e., $d > 0.95$). And the Bell nonlocality disappear iff $d > 0.5$, that is, this correlation is only locality. Intuitively, the quantum steering and Bell nonlocality are very stronger iff their state is in a maximally entangled evolution one, meanwhile, decoherence strength should be small enough. Besides, in BF channel, we can obtain that quantum steering and Bell nonlocality are symmetrical about $p = 0.5$, and all steerable states can violate the Bell-CHSH inequality (see Fig. 2 (c) and (d)).

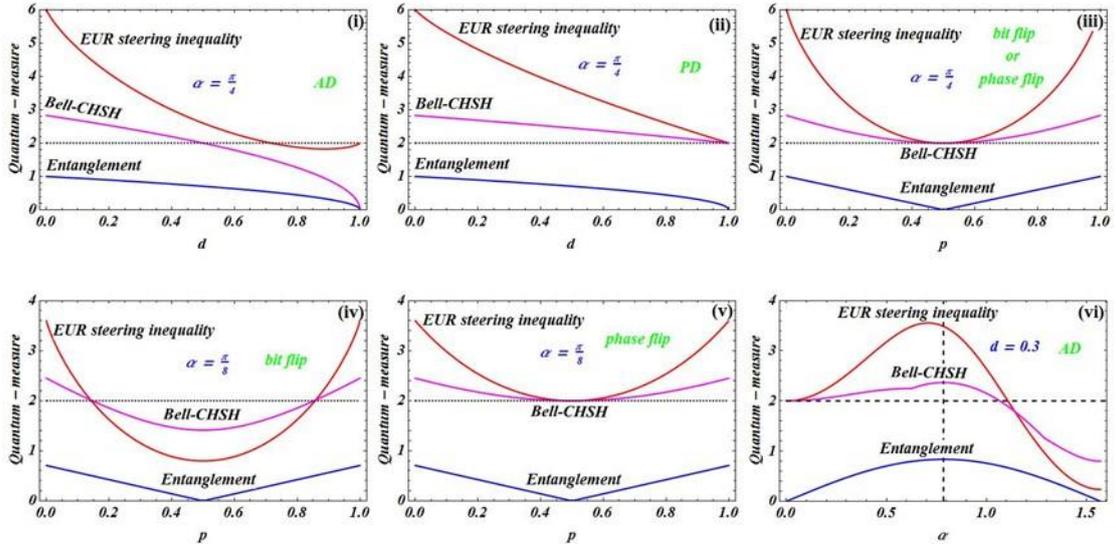

**Fig. 3.** (Color online) A variety of quantum-measure (*EUR steering inequality*, Bell-CHSH inequality and entanglement) as function of decoherence strength $d$, $p$ for the maximally entangled state $\alpha = \pi/4$ (shown in **(i)**, **(ii)**

and **(iii)**). Quantum-measure as function of decoherence strength $p$ for $\alpha = \pi/8$ (shown in **(iv)** and **(v)**).**(vi)** The quantum-measure as function of state parameters $\alpha$ for $d = 0.3$.

Subsequently, the relationships among three quantum measures: entanglement, quantum steering and nonlocality in the different quantum channels are shown in Fig. 3. From the figure, one can obtain that every maximally entangled evolution state is maximally steerable state. Some EESs are unsteerable and some steerable states will not obey Bell nonlocality. However, in PD and PF channels, all EESs are steerable and satisfy Bell nonlocality. In addition, all steerable states can violate the Bell-CHSH inequality, but some EESs cannot give rise to steering in BF channel. Apart from that the steerability of the initial entangled state is destroyed by decoherence, quantum steering experiences a recovery with the increase of decoherence strength in BF and PF channels. Moreover, all EESs can violate *EUR steering inequality* and satisfy Bell nonlocality in PD and PF channels (shown as Fig. 3 (ii), (iii) and (vi)). In AD channel, we can find that the symmetry of quantum steering (or nonlocality) of the initial state is destroyed, but that does not for quantum entanglement (see Fig. 3 (vi)).

**Alice and Bob share an entangled mixed state.** Considering the mixed state[35]

$$\rho(v) = v|\psi\rangle\langle\psi| + (1-v)|\varphi\rangle\langle\varphi|, \tag{7}$$

where $|\psi\rangle = \dfrac{|00\rangle + |11\rangle}{\sqrt{2}}$ and $|\varphi\rangle = \dfrac{|01\rangle + |10\rangle}{\sqrt{2}}$. It is entangled when $v \in [0, 1/2) \cup (1/2, 1]$. Then, we still consider previous physical case as shown in Fig. 1 (a). For convenience, we display the corresponding each parameter expression of two-qubit EESs in Bloch decomposition in Table. 3.

|       | AD                    | PD                    | BF            |
|-------|-----------------------|-----------------------|---------------|
| $c_1$ | $\sqrt{1-d}$          | $\sqrt{1-d}$          | 1             |
| $c_2$ | $(1-2v)\sqrt{1-d}$    | $(1-2v)\sqrt{1-d}$    | $(2v-1)(1-2p)$ |
| $c_3$ | $(2v-1)(1-d)$         | $2v-1$                | $(2v-1)(2p-1)$ |
| $r$   | 0                     | 0                     | 0             |
| $s$   | $d$                   | 0                     | 0             |

**Table. 3.** The corresponding expressions of each parameter of two-qubit EESs in Bloch decomposition are given in the different noisy channels when Alice and Bob initially share an entangled mixed state.

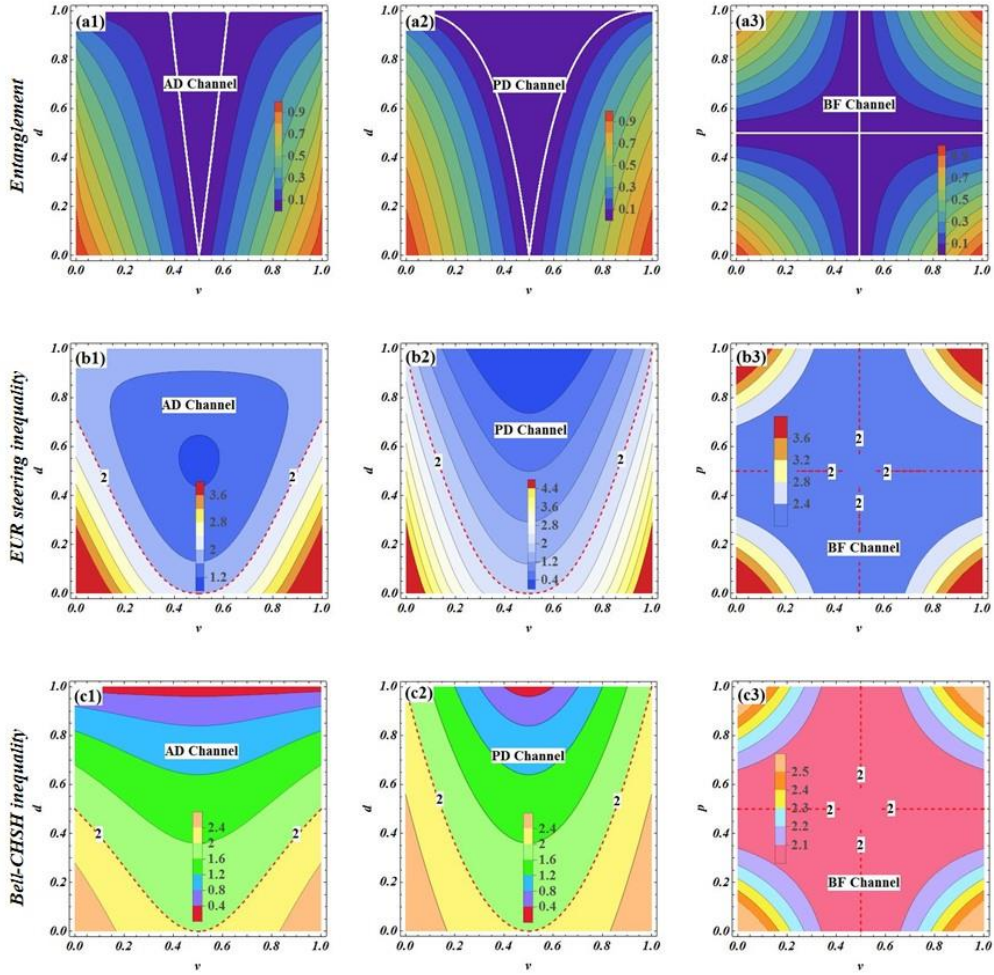

**Fig. 4.** (Color online) Contour plot of entanglement (concurrence), *EUR steering inequality* and Bell-CHSH inequality versus decoherence strength *d* and states parameters *v*, AD channel shown in **(a1)**, **(b1)** and **(c1)**; PD channel shown in **(a2)**, **(b2)** and **(c2)**, respectively. The bottom of the red dotted line denotes the steering (see **(b1)** and **(b2)**) and Bell nonlocality (see **(c1)** and **(c2)**). For BF channel, contour plot of entanglement (concurrence), *EUR steering inequality* versus decoherence strength *p* and states parameters *v* in **(a3)**, **(b3)** and **(c3)**, respectively, when initial state is an entangled mixed state.

In order to better comprehend the relationship among entanglement, quantum steering and nonlocality in the different types of noises, we draw the counterpart contour plots in Fig. 4. From the figure, we can obtain that all EESs' entanglement; steering and nonlocality will experience a "sudden death". Some EESs are unsteerable and some steerable states do not obey Bell nonlocality in AD channel. In addition, some results are not the same as the above case (the initial state is an entangled pure state). We find that all EESs can be employed to realize steering and satisfy Bell nonlocality in BF channel. However, in PD channel, all steerable states can violate the Bell-CHSH inequality, but some EESs cannot violate *EUR steering inequality*. Furthermore, in AD channel, decoherence can destroy the steerability of the initial state, and until the EESs cannot steer ($d > 0.7$), and the Bell nonlocality is absent iff $d > 0.5$. Moreover, quantum steering experiences a recovery with increasing state parameters $v$ when decoherence strength is a fixed value in any noisy channel.

Via the analysis, one can conclude that the steerability of entangled mixed states is weaker than the steerability of entangled pure states, and the steerability of state is associated with the interaction between quantum systems and quantum channels. Furthermore, the steering behaves sometimes like the nonlocality and sometimes like the entanglement. That is, quantum steering is an intermediate form of quantum correlation between entanglement and nonlocality.

**Conclusions**

To conclude, we analytically derive the performance of quantum steering, nonlocality and entanglement, and discuss the relationship among them in structured reservoirs for two different types of initial states: entangled pure state and entangled mixed state. Our results indicate that the steerability of entangled pure states is stronger than that of entangled mixed states, and entangled pure states and the maximally EESs are steerable. Not every entangled evolution state is steerable and some steerable states cannot violate Bell-CHSH inequality. In other words, if an entangled state shared by Alice and Bob is steerable, when the state suffers from the reservoirs, the state may be unsteerable, meanwhile, the Bell nonlocality may be absent.

Importantly, we find that all EESs can violate *EUR steering inequality* and Bell-CHSH inequality in PD and PF channels when they initially share an entangled pure state. In BF channel, all steerable states can satisfy Bell nonlocality, but some EESs are unsteerable. However, when they initially share an entangled mixed state, all EESs can be employed to realize steering and can lead to Bell nonlocality in BF channel. Moreover, decoherence can effectively induce the degradation of quantum steering, nonlocality and entanglement. However, these quantum correlations experience a recovery with the increase of decoherence strength in BF and PF channels. Therefore, we could say, the steerability of state is associated with the interaction between quantum systems and external noises.

# Methods

## Quantum entanglement, nonlocality and steering of two-qubit *X*-state

We first introduce the form of two-qubit *X*-state. The *X*-shaped states, which are represented in the orthonormal basis $\{|00\rangle, |01\rangle, |10\rangle, |11\rangle\}$ as

$$\rho^X = \begin{pmatrix} \rho_{11} & 0 & 0 & \rho_{14} \\ 0 & \rho_{22} & \rho_{23} & 0 \\ 0 & \rho_{23} & \rho_{33} & 0 \\ \rho_{14} & 0 & 0 & \rho_{44} \end{pmatrix}, \tag{8}$$

where $\rho_{ij}(i, j = 1, 2, 3, 4)$ are all real parameters. As is well known, the degree of entanglement for bipartite states can be quantified conveniently by concurrence. Hence, we chose concurrence as entanglement measurement. The concurrence is defined as[48, 49]

$$C = \max\{0, \sqrt{\lambda_1} - \sqrt{\lambda_2} - \sqrt{\lambda_3} - \sqrt{\lambda_4}\}, \quad \lambda_1 \geq \lambda_2 \geq \lambda_3 \geq \lambda_4 \geq 0, \tag{9}$$

where $\lambda_i (i = 1, 2, 3, 4)$ are the eigenvalues of the matrix $R = \rho(\sigma_y \otimes \sigma_y)\rho^*(\sigma_y \otimes \sigma_y)$. The density matrix is *X*-structure, there is a reduced form for concurrence shown as[50]

$$C = 2\max\{0, |\rho_{14}| - \sqrt{\rho_{22}\rho_{33}}, |\rho_{23}| - \sqrt{\rho_{11}\rho_{44}}\}, \tag{10}$$

where $\rho_{ij}$ are the elements of the matrix $\rho^X$. Thus, employing Eq. (10), we can obtain the expressions of concurrence in the different quantum channels $C_{AD} = C_{PD} = \sqrt{1-d}\sin(2\alpha)$, $C_{PF} = C_{BF} = |2p-1|\sin(2\alpha)$, respectively, when the initial state is an entangled pure state (1).

While initial state is an entangled mixed state (7), the concurrence in the different quantum channels can be expression as

$$C_{AD}^M = \max\{0, \sqrt{1-d}(1-v) - \sqrt{v(1-d)(v+d(1-v))}, \\ v\sqrt{1-d} - \sqrt{(1-d)(1-v)((1-v)+vd)}\}, \tag{11}$$

$$C_{PD}^M = \max\{0, \sqrt{1-d}(1-v) - v, v\sqrt{1-d} - (1-v)\}, \tag{12}$$

$$C_{BF}^M = \max\{0, 2(1-v+p(2v-1)) - 2\sqrt{\Theta}/2\sqrt{2}, \\ (p+v-2pv) - \sqrt{(pv+(p-1)(v-1))((p-1)(v-1)+pv)}\}, \tag{13}$$

with

$$\Theta = p^2 + 2(3-4p)pv + (1+8(p-1)p)v^2 + (p-v)^2, \qquad (14)$$

respectively. Then, by employing appropriate local unitary transformations, one can rewrite the state $\rho^X$ of Eq. (8) in Bloch decomposition

$$\rho^X = \frac{1}{4}\left(I^A \otimes I^B + \vec{r}\cdot\sigma^A \otimes I^B + I^A \otimes \vec{s}\cdot\sigma^B + \sum_{i=1}^{3} c_i \sigma_i^A \otimes \sigma_i^B\right), \qquad (15)$$

where $\vec{r}$ and $\vec{s}$ are Bloch vectors, and $\{\sigma_i\}_{i=1}^{3}$ are standard Pauli matrices. If $\vec{r}=\vec{s}=\mathbf{0}$, $\rho^X$ is the a two-qubit Bell-diagonal state. Assume that Bloch vectors are in the $z$ direction, that is, $\vec{r}=(0,\ 0,\ r)$, $\vec{s}=(0,\ 0,\ s)$, the density matrix of $\rho^X$ in Eq. (15) has the following form

$$\rho^X = \frac{1}{4}\begin{pmatrix} 1+c_3+s+r & 0 & 0 & c_1-c_2 \\ 0 & 1-c_3+r-s & c_1+c_2 & 0 \\ 0 & c_1+c_2 & 1-c_3-r+s & 0 \\ c_1-c_2 & 0 & 0 & 1+c_3-r-s \end{pmatrix}, \qquad (16)$$

with

$$\begin{aligned} c_1 &= 2(\rho_{23}+\rho_{14}), \\ c_2 &= 2(\rho_{23}-\rho_{14}), \\ c_3 &= \rho_{11}-\rho_{22}-\rho_{33}+\rho_{44}, \\ r &= \rho_{11}+\rho_{22}-\rho_{33}-\rho_{44}, \\ s &= \rho_{11}-\rho_{22}+\rho_{33}-\rho_{44}. \end{aligned} \qquad (17)$$

According to the Horodecki criterion[2, 3], $B = 2\sqrt{\max_{i<j}(\mu_i+\mu_j)}$ with $i, j = 1,\ 2,\ 3$. The three eigenvalues $\mu_i$ of $U = T^T T$ for X-state are

$$\mu_1 = 4(|\rho_{14}|+|\rho_{23}|)^2,\ \mu_2 = 4(|\rho_{14}|-|\rho_{23}|)^2,\ \mu_3 = (\rho_{11}-\rho_{22}-\rho_{33}+\rho_{44})^2. \qquad (18)$$

It is easy to see that $\mu_1$ is always larger than $\mu_2$, and thus the Bell Clauser-Horne-Shimony-Holt (Bell-CHSH) inequality maximum violation of X-state is[51-53]

$$B = 2\max\{B_1,\ B_2\},\ B_1 = \sqrt{\mu_1+\mu_2},\ B_2 = \sqrt{\mu_1+\mu_3}. \qquad (19)$$

When Alice and Bob initially share an entangled mixed state (7), we can obtain the expressions of Bell-CHSH inequality in the different quantum channels

$$B_{AD}^M = 2\sqrt{(1-d)(1+(1-2v)^2)},\ B_{PD}^M = 2\sqrt{1-d+(1-2v)^2},\ B_{BF}^M = 2\sqrt{1+(1-2v)^2(1-2p)^2}, \qquad (20)$$

respectively. Subsequently, depending upon *EUR steering inequality*'s definition in Eq. (6),

employing the X-state $\rho^X$ in Eq. (16), we can obtain the expression of *EUR steering inequality* for the general bipartite X-state by using Pauli *X*, *Y*, and *Z* measurements on each side

$$\sum_{i=1,2}\left[(1+c_i)\log(1+c_i)+(1-c_i)\log(1-c_i)\right]-(1+r)\log(1+r)-(1-r)\log(1-r)$$
$$+\frac{1}{2}\left[(1+c_3+r+s)\log(1+c_3+r+s)+(1+c_3-r-s)\log(1+c_3-r-s)\right. \tag{21}$$
$$\left.+(1-c_3-r+s)\log(1-c_3-r+s)+(1-c_3+r-s)\log(1-c_3+r-s)\right]\leq 2.$$

If $r=s=0$, the bipartite X-state will become the Bell-diagonal states. The Eq. (21) is simplified into[30] $\sum_{i=1,2,3}(1+c_i)\log(1+c_i)+(1-c_i)\log(1-c_i)\leq 2$. As an explanation, employing measurement in the Pauli *X* bases on each side, the four eigenvalues of the bipartite X-state $\rho^x_{AB}$ are $\{\lambda_{x1}=\lambda_{x2}=(1-c_1)/4,\ \lambda_{x3}=\lambda_{x4}=(1+c_1)/4\}$, and the two eigenvalues of the reduced state $\rho^x_A=Tr_B[\rho^x_{AB}]$ are $\{\lambda^A_{x1}=\lambda^A_{x2}=1/2\}$. In the same way, we can obtain that the eigenvalues of the other two bipartite X-state are $\{\lambda_{y1}=\lambda_{y2}=(1-c_2)/4,\ \lambda_{y3}=\lambda_{y4}=(1+c_2)/4\}$ and $\{\lambda_{z1}=(1-c_3+r-s)/4,\ \lambda_{z2}=(1-c_3-r+s)/4,\ \lambda_{z3}=(1+c_3+r+s)/4,\ \lambda_{z4}=(1+c_3-r-s)/4\}$ by using Pauli *Y* and *Z* measurements on each side, respectively. The corresponding the eigenvalues of the reduced states $\rho^y_A$, $\rho^z_A$ are $\{\lambda^A_{y1}=\lambda^A_{y2}=1/2\}$ and $\{\lambda^A_{z1}=(1-r)/2,\ \lambda^A_{z2}=(1+r)/2\}$, respectively. Then, it is straightforward to insert all above eigenvalues into Eq. (6), we can obtain the expression of *EUR steering inequality*. Finally, it is straightforward to insert each parameter of Table. 2 and Table. 3 into Eqs. (19) and (21), resulting in the analytical expressions of Bell-CHSH inequality and *EUR steering inequality*.

## Acknowledgments

This work was supported by the National Science Foundation of China under Grant Nos. 11575001, 61275119, 61601002 and 11605028, Anhui Provincial Natural Science Foundation (Grant No. 1508085QF139).

## Author Contributions

W.-Y. Sun, D. Wang and J.-D. Shi performed the calculations and wrote the paper with advice from L. Ye; W.-Y. Sun and L. Ye devised the theoretical scheme and provided the theoretical analysis; and all authors reviewed the manuscript and agreed with the submission.

## Additional Information

**Competing financial interests:** The authors declare no competing financial interests.